\documentclass[a4paper,twocolumn,english,prb, eqsecnum]{revtex4}
\usepackage{times}
\usepackage[T1]{fontenc}
\usepackage[latin1]{inputenc}
\usepackage{amsmath}
\usepackage{amssymb}

\makeatletter

\date{\today}
\usepackage{ae,aecompl}

\usepackage{bm}

\usepackage{babel}
\makeatother
\begin{document}

\title{Gaussian phase-space representations for fermions}

\author{J. F. Corney and P. D. Drummond}

\affiliation{ARC Centre of Excellence for Quantum-Atom Optics, University of Queensland,
Brisbane 4072, Queensland, Australia.}

\begin{abstract}
We introduce a positive phase-space representation for fermions, using
the most general possible multi-mode Gaussian operator basis. The
representation generalizes previous bosonic quantum phase-space methods
to Fermi systems. We derive equivalences between quantum and stochastic
moments, as well as operator correspondences that map quantum operator
evolution onto stochastic processes in phase space. The representation
thus enables first-principles quantum dynamical or equilibrium calculations
in many-body Fermi systems. Potential applications are to strongly
interacting and correlated Fermi gases, including coherent behaviour
in open systems and nanostructures described by master equations.
Examples of an ideal gas and the Hubbard model are given, as well
as a generic open system, in order to illustrate these ideas.
\end{abstract}
\maketitle

\section{Introduction}

The study of strongly correlated Fermi gases is one of the most active
areas in modern condensed matter and AMO physics. In quantum degenerate
electron gases, improvements in condensed matter materials have led
to sophisticated experiments, typically in reduced dimensional environments.
Many interesting quantum phenomena are observed in these systems,
including such features as the quantum Hall effect\cite{Quanthall},
metal-insulator phase-transitions\cite{Metal-Insulator}, high T-c
superconductors\cite{HighTc}, and single electron gates in nanostructures\cite{nano}. 

Recently, pioneering experiments in strongly-interacting ultra-cold
Fermi gases have opened up novel experiments of unprecedented simplicity
and precision, both in the BEC-BCS cross-over regime\cite{Experiments},
and in lattices\cite{LatticeFermiGas}. The underlying atomic interactions
are extremely well-understood, and the dynamics, interactions and
geometry are all highly adaptable. Measurement techniques are also
rapidly improving, with direct measurements of collective modes\cite{Collective},
thermodynamic properties\cite{Thermo}, vortices\cite{Vortex}, and
even momentum correlations being recently reported\cite{Correlation}. 

This situation provides a substantial opportunity to develop and test
first-principles theoretical methods for the investigation of correlations
and dynamical effects in quantum degenerate Fermi gases. To this end,
we introduce a generalised phase-space representation for correlated
fermionic systems. The representation is based on a Gaussian operator
basis for fermionic density operators. Like the analogous basis for
bosons\cite{Gauss:Bosons}, the fermionic operator basis enables the
representation of arbitrary physical density operators as a positive
distribution over a phase space. This representation allows quantum
evolution, either in real time or in inverse temperature, to be viewed
as a stochastic evolution of covariances or Green's functions.

Phase-space methods based on coherent states\cite{Schrod} have long
been used for bosonic systems, with great success. These approaches
include the Wigner function\cite{Wig-Wigner}, the Q-function\cite{Hus-Q},
as well as the well-known Glauber-Sudarshan $P$-function\cite{Gla-P},
and its generalizations\cite{CG-Q,Agarwal}. The early methods based
on classical phase spaces were later generalized to give the positive-P
distribution\cite{positiveP1}, which has proved a successful way
to simulate quantum many-body systems from first principles\cite{ICOLSProceedings}.
This method reduces quantum dynamics to the time evolution of a positive
distribution on an over-complete basis set of coherent state projection
operators, which are special cases of the bosonic Gaussian operators.
Applications have been to quantum statistics of lasers\cite{Louisell},
superfluorescence\cite{Super}, parametric amplifiers\cite{positiveP1,QuantNoise},
quantum solitons\cite{Soliton}, as well as quantum dynamics\cite{BEC}
and thermal correlations\cite{BEC1d} in Bose-Einstein condensates.

\emph{Fermionic} phase-space representations are relevant to a long-standing
problem in theoretical physics, which is the sign problem that occurs
in many-body fermionic wave-functions\cite{Ceperley,Linden92,Santos03}.
There are many different approximate techniques that can be used,
but the intention of this paper is to establish fundamentally exact
procedures to treat the Fermi sign problem. As shown in \cite{GaussianQMC},
the Gaussian method can be applied to the difficult case of the repulsive
Hubbard model\cite{Hubbard63}. Here we concentrate on the foundational
issues of the Gaussian representation method, presenting the general
identities required to apply the method to a wide range of problems
in fermionic many-body physics, including both ultra-cold atomic and
condensed matter systems.

To proceed, we make use of three important results, proved elsewhere\cite{CorneyDrummond2005A}: 

\begin{itemize}
\item the Gaussian fermion operators form a complete basis for any physical
density operator, 
\item the distribution can always be chosen positive, and
\item there are mappings to a second-order differential form for all two-body
operators. 
\end{itemize}
From these properties, we show that positive-definite Fokker-Planck
equations exist for many-body fermionic systems, provided that the
distribution tails remain sufficiently bounded. Such Fokker-Planck
equations enable first-principles stochastic simulation methods, either
in real time or at finite temperature. As is usual in such methods,
care must be taken with sampling errors and boundary terms due to
the distribution tails. Due to the non-uniqueness of the representation,
there is a type of gauge freedom in the choice of stochastic equation.
We show how this stochastic gauge freedom, which has been successfully
used to remove boundary terms in bosonic representations\cite{DeuarDrummond02},
can in principle also be used here. 

Representations for fermionic density operators were introduced by
Cahill and Glauber\cite{CahillGlauber99} using fermionic coherent
states\cite{FermiCoherentState}. These provide a means of defining
quasi-probabilities for fermionic states analogous to the well-known
bosonic distributions\cite{CahillGlauber99,PlimakCollettOlsen01}.
However, the resulting quasi-probabilities are functions of non-commuting
Grassmann variables, and are thus not directly computationally accessible.
Nevertheless, fermion coherent states and Grassmann algebra are useful
for deriving analytical results in Fermi systems. 

The Gaussian method introduced here overcomes the problems inherent
in using Grassmann algebra variables. The Gaussian expansion utilises
an operator basis constructed from pairs \emph{}of operators, instead
of a state-vector basis. Because pairs of fermion operators obey commutation
relations rather than anti-commutation relations, a natural solution
of the anticommutation problem is achieved. The resulting phase space
thus exists on a domain of commuting c-numbers, rather than anti-commuting
Grassmann variables. Furthermore, the phase-space equations obviate
the need to evaluate large determinants in simulations. This method
substantially generalizes and extends earlier phase-space techniques
used in quantum optics to treat electronic transitions in atoms\cite{Louisell,Oldfermi}.
It is different to auxiliary field quantum Monte Carlo methods\cite{RomboutsHeyde03}
in condensed matter theory, which use Gaussian operators, but involve
path integrals rather than positive expansions of the density matrix.

We begin in Sec.\ref{sec:Gaussian-operators} by defining the Gaussian
operator basis on which the representation is based, and introducing
some convenient notations. In Sec.\ref{sec:Gaussian-Representation},
we define the Gaussian representation as an expansion in Gaussian
operators, and then show how the representation establishes a novel
class of exact Monte-Carlo type methods for simulating the real-time
dynamics or finite-temperature equilibrium of a quantum system. We
show how to map quantum operator evolution onto a set of stochastic
(real or complex) differential equations, and give the correspondences
necessary to calculate physical moments. 

Finally, in Sec.~\ref{sec:Examples}, we give examples of the application
of the method. These are intended to be illustrative rather than exhaustive,
and further examples and applications in greater detail will be given
elsewhere. In particular, we note that any nonlinear application requires
a careful analysis of the issues of sampling error and boundary term
behaviour. For simplicity, we focus on the ideal Fermi gas, a generic
open system master equation and the finite temperature Hubbard model,
as well as showing how to apply gauges to modify the drift evolution.

\section{Gaussian operators}

\label{sec:Gaussian-operators}

Before discussing the Gaussian representation, we first introduce
the fermion operators on which it is based. Fermionic Gaussian operators
are defined as exponentials of quadratic forms in the Fermi annihilation
or creation operators. This simple definition encompasses a wide range
of physical applicability. Obviously, it includes the well-known thermal
density matrices of the free field. Since the definition includes
quadratic forms involving pairs of annihilation or creation operators,
it also encompasses the pure-state density matrices that correspond
to the BCS states used in superconductivity.

A more subtle issue is that the definition is not restricted to Hermitian
operators. This has the advantage of leading to completeness properties
that are much stronger than if the definition were restricted to only
Hermitian operators. Some of these issues are discussed elsewhere,
in a more formal derivation of the mathematical properties of the
Gaussian operators\cite{CorneyDrummond2005A}.

\subsection{Notation}

Before giving mathematical results, we summarize the notation that
will be used. We can decompose a given fermionic system into a set
of $M$ orthogonal single-particle modes, or orbitals. With each of
these modes, we associate creation and annihilation operators $\widehat{b}_{j}^{\dagger}$
and $\widehat{b}_{j}$, with anticommutation relations \begin{eqnarray}
\left[\widehat{b}_{k},\widehat{b}_{j}^{\dagger}\right]_{+} & = & \delta_{kj}\nonumber \\
\left[\widehat{b}_{k},\widehat{b}_{j}\right]_{+} & = & 0\,\,,\label{eq:anticommute}\end{eqnarray}
 where $j,k=1...M$. Thus, $\widehat{\bm b}$ is a column vector of
the $M$ annihilation operators, and $\widehat{\bm b}^{\dagger}$
is a row vector of the corresponding creation operators.

For products of operators, we make use of normal and antinormal ordering
concepts. Normal ordering, denoted by $:\cdots:$ , is defined as
in the bosonic case, with all annihilation operators to the right
of the creation operators, except that each pairwise reordering involved
induces a sign change, e.~g.~$:\widehat{b}_{i}\widehat{b}_{j}^{\dagger}:\,=-\widehat{b}_{j}^{\dagger}\widehat{b}_{i}\,\,$.
The sign changes are necessary so that the anticommuting natures of
the Fermi operators can be accommodated without ambiguity. 

To enable the general Gaussian operator to be written in a compact
form, we use an extended-vector notation: \begin{equation}
\underline{\widehat{b}}=\left(\begin{array}{c}
\widehat{\bm b}\\
\widehat{\bm b}^{\dagger T}\end{array}\right)\,\,,\end{equation}
 is defined as an extended column vector of all $2M$ operators, with
an adjoint row vector defined as \begin{equation}
\underline{\widehat{b}}^{\dagger}=(\widehat{\bm b}^{\dagger},\widehat{\bm b}^{T})\,\,.\end{equation}
 Throughout the paper, we print vectors of length $M$ and $M\times M$
matrices in bold type, and index them where necessary with Latin indices:
$j=1,...,M$. Vectors of length $2M$ we denote with an underline,
while $2M\times2M$ matrices are indicated by a double underline.
These extended vectors and matrices are indexed where necessary with
Greek indices: $\mu=1,...,2M$. For further examples of this notation,
see \cite{CorneyDrummond2005A,Gauss:Bosons}. More general kinds of
vectors are denoted with an arrow notation: $\overrightarrow{\lambda}$.

\subsection{Definition of the Gaussian operator}

We define a Gaussian operator to be any normally ordered, Gaussian
form of annihilation and creation operators. Like a complex number
Gaussian, the operator Gaussian is an exponential of a quadratic form,
with the exponential defined by its series representation. The most
general Gaussian form is a cumbersome object to manipulate, unless
products of odd numbers of operators are excluded. Fortunately, restricting
the set of Gaussians to those containing only even products can be
physically justified on the basis of superselection rules for fermions.
Because it is constructed from pairs of operators, this type of Gaussian
operator contains no Grassmann variables.

With the extended-vector notation, we can write any general Gaussian
operator $\widehat{\Lambda}$ as:\begin{equation}
\widehat{\Lambda}(\overrightarrow{\lambda})=\Omega\frac{1}{{\cal N}}:\exp\left[-\underline{\widehat{b}}^{\dagger}\,\underline{\underline{\Sigma}}\,\underline{\widehat{b}}/2\right]:\,,\label{eq:Gaussbasis}\end{equation}
where $\Omega$ is an amplitude, ${\cal N}$ is a normalizing factor
defined so that $\mathrm{Tr}\left[\widehat{\Lambda}(\overrightarrow{\lambda})\right]=\Omega$,
and $\underline{\underline{\Sigma}}$ is a $2M\times2M$ complex matrix.
For later identification with physical observables, it proves useful
to write $\underline{\underline{\Sigma}}$ in the form:\begin{equation}
\underline{\underline{\Sigma}}=\left(\underline{\underline{\sigma}}^{-1}-2\underline{\underline{I}}\right),\end{equation}
 where $\underline{\underline{\sigma}}$ is a generalised covariance
matrix and $\underline{\underline{I}}$ is the constant matrix is
defined as\begin{eqnarray}
\underline{\underline{I}} & = & \left[\begin{array}{cc}
-\mathbf{I} & \mathbf{0}\\
\mathbf{0} & \mathbf{I}\end{array}\right]\,\,,\end{eqnarray}
It is convenient to introduce complex $M\times M$ matrices $\mathbf{n}$
and $\tilde{\mathbf{n}}=\mathbf{I}-\mathbf{n}$ which, as we show
later, correspond to normal Green's function for particles and holes
respectively, and two independent antisymmetric complex $M\times M$
matrices $\mathbf{m}$ and $\mathbf{m}^{+}$ that correspond to anomalous
Green's functions. These are related to the covariance matrix $\underline{\underline{\sigma}}$
by \begin{eqnarray}
\left[\begin{array}{cc}
-\tilde{\mathbf{n}}^{T} & \mathbf{m}\\
\mathbf{m}^{+} & \tilde{\mathbf{n}}\end{array}\right]=\left[\begin{array}{cc}
\mathbf{n^{T}}-\mathbf{I} & \mathbf{m}\\
\mathbf{m}^{+} & \mathbf{I}-\mathbf{n}\end{array}\right] & = & \frac{1}{2}\left(\underline{\underline{\sigma}}-\underline{\underline{\sigma}}^{+}\right),\nonumber \\
\label{eq:sigma}\end{eqnarray}
where `$+$' denotes a generalised transpose operation defined by

\begin{equation}
\left[\begin{array}{cc}
\mathbf{a} & \mathbf{b}\\
\mathbf{c} & \mathbf{d}\end{array}\right]^{+}\equiv\left[\begin{array}{cc}
\mathbf{d} & \mathbf{c}\\
\mathbf{b} & \mathbf{a}\end{array}\right]^{T}\,\,.\label{eq:sym}\end{equation}
 Any Gaussian operator can always be written with a covariance matrix
that has the antisymmetry $\underline{\underline{\sigma}}=-\underline{\underline{\sigma}}^{+}$.

The Gaussian operators are defined here in terms of the full $1+p=1+4M^{2}$
amplitude plus covariance matrix components. Alternatively, we can
include just the parameters $\overrightarrow{\lambda}$ that lead
to distinguishable Gaussians, where\begin{equation}
\overrightarrow{\lambda}=(\Omega,\mathbf{n},\mathbf{m},\mathbf{m}^{+})\,\,,\label{eq:lambda}\end{equation}
giving only $1+p=1+M(2M-1)$ parameters. We will index over the phase-space
variables with the notation $\lambda_{a}$, $a=0,\ldots p$. For simplicity,
we generally deal with the full, unconstrained $\underline{\underline{\sigma}}$
matrices.

The normalisation ${\cal N}$ contains a Pfaffian whose square is
equal to the determinant of the matrix. We will show that ${\cal N}$
does not appear explicitly in later results. The additional variable
$\Omega$ plays the role of a weighting factor in the expansion. This
allows us to represent unnormalised density operators like $\exp(-\beta\widehat{H}$),
and to introduce stochastic gauges that change these relative weighting
factors in order to stabilize trajectories.

\subsection{Moments}

\label{sub:Moments}

Just as with classical Gaussian forms, these generalised fermionic
Gaussians are completely characterised by their first order moments
(to within a weight factor): \begin{eqnarray}
\mathrm{Tr}\left[\widehat{b}_{i}\widehat{b}_{j}\widehat{\Lambda}\right] & = & \Omega m_{ij},\nonumber \\
\mathrm{Tr}\left[\widehat{b}_{i}^{\dagger}\widehat{b}_{j}\widehat{\Lambda}\right] & = & \Omega n_{ij}\,,\nonumber \\
\mathrm{Tr}\left[\widehat{b}_{j}\widehat{b}_{i}^{\dagger}\widehat{\Lambda}\right] & = & \Omega\tilde{n}_{ij}\,,\nonumber \\
\mathrm{Tr}\left[\widehat{b}_{i}^{\dagger}\widehat{b}_{j}^{\dagger}\widehat{\Lambda}\right] & = & \Omega m_{ij}^{+}\,.\label{moments}\end{eqnarray}
If the Gaussian operator happens to be a physical density matrix,
these quantities correspond to the first-order correlations or Green's
functions. Thus, in many-body terminology, $\mathbf{n}$ and $\tilde{\mathbf{n}}$
are the normal Green's functions of particles and holes, respectively,
and $\mathbf{m}$ and $\mathbf{m}^{+}$ are anomalous Green's functions.
From this we see that, for the subset of Gaussians that are physical
density matrices, we must have that $\mathbf{m}^{\dagger}$=$\mathbf{m^{+}}$,
and $\mathbf{n}^{\dagger}$=$\mathbf{n}$. Furthermore, $\mathbf{n}$
and $\tilde{\mathbf{n}}$ must be positive semi-definite (because
$0\le\left\langle \widehat{n}_{jj}\right\rangle \le1$) .

More generally, the phase-space function $O(\overrightarrow{\lambda})$
corresponding to the normally ordered operator $\widehat{O}$ is defined
as a phase-space correspondence, according to:\begin{equation}
O(\overrightarrow{\lambda})\equiv\left\langle \widehat{O}\right\rangle _{\overrightarrow{\lambda}}\equiv{\rm Tr}\left(\widehat{O}\,\widehat{\Lambda}(\overrightarrow{\lambda})\right)/\Omega\,\,.\label{eq:correspondence}\end{equation}
For higher-order moments, a form of Wick's theorem applies to any
normally ordered product. One simply writes down the sum of all distinct
factorisations into pairs, with a minus sign in front of any product
that is an odd permutation of the original form. The term distinct
factorization means that neither permutation of pair ordering nor
re-ordering inside a pair is regarded as significant, since these
do not change the result. Thus an $N$-th order correlation (expectation
value of a product of $2N$ operators), is the sum of $2N!/(2^{N}N!)$
distinct terms, as follows,\begin{eqnarray}
\left\langle :\widehat{b}_{\mu_{1}}....\widehat{b}_{\mu_{2N}}:\right\rangle _{\overrightarrow{\lambda}} & = & \sum_{P}(-1)^{P}\left\langle :\widehat{b}_{\mu_{P(1)}}\widehat{b}_{\mu_{P(2)}}:\right\rangle _{\overrightarrow{\lambda}}\times...\nonumber \\
 &  & \times\left\langle :\widehat{b}_{\mu_{P(2N-1)}}\widehat{b}_{\mu_{P(2N)}}:\right\rangle _{\overrightarrow{\lambda}}.\label{eq:Wick}\end{eqnarray}
Here the sum is over all $2N!/(2^{N}N!)$ distinct pair permutations
$P(1),...,P(2N)$ of $1,...,2N$, and where $(-1)^{P}$ is the parity
of the permutation (i.e. the number of pair-wise transpositions required
to perform the permutation).

Thus, for example, the second-order number correlation moment is:\begin{equation}
\left\langle \widehat{b}_{i}^{\dagger}\widehat{b}_{j}^{\dagger}\widehat{b}_{j}\widehat{b}_{i}\right\rangle _{\overrightarrow{\lambda}}=n_{ii}n_{jj}-n_{ij}n_{ji}+m_{ij}^{+}m_{ji}\,\,.\end{equation}

\subsection{Generalised thermal states}

An important subset of the Gaussian operators is the set of generalized
thermal operators, for which $\mathbf{m}=\mathbf{m}^{+}=\mathbf{0}$.
These include the canonical density matrices for free Fermi gases
in the case that $\mathbf{n}$, and $\tilde{\mathbf{n}}$ are each
Hermitian and positive definite. More generally, however, we do not
require $\mathbf{n}$ to be Hermitian. In all cases, the generalized
thermal operators in normally ordered Gaussian form can be written
most directly in terms of the hole population, $\tilde{\mathbf{n}}=\mathbf{I}-\mathbf{n}$:

\begin{eqnarray}
\widehat{\Lambda}(\overrightarrow{\lambda}) & = & \Omega\det\left[\tilde{\mathbf{n}}\right]:\exp\left[\widehat{\mathbf{b}}^{\dagger}\left(\tilde{\mathbf{n}}^{-1}-2\mathbf{I}\right)^{T}\widehat{\mathbf{b}}\right]:\,\,.\nonumber \\
\label{eq:ThermalGaussian}\end{eqnarray}
Of course, there is a symmetry here: in an antinormally ordered Gaussian,
the role of $\widehat{\mathbf{b}}^{\dagger}$ and $\widehat{\mathbf{b}}$
is reversed, and consequently so is the role of $\mathbf{n}$ and
$\tilde{\mathbf{n}}$. Our choice of normal ordering is in fact arbitrary
from a physical point of view, and antinormal ordering would also
serve our purpose equally well, provided all the identities were redefined.

By comparison, the usual canonical form of the fermionic thermal state
with a diagonal Hamiltonian $H=\widehat{\mathbf{b}}^{\dagger}\mathbf{{\bf \omega}}\widehat{\mathbf{b}}$
and a chemical potential $\mu,$ is an unordered form, namely:\begin{equation}
\widehat{\rho}(\tau)=\exp\left[\tau\widehat{\mathbf{b}}^{\dagger}(\mu-\mathbf{\omega})\widehat{\mathbf{b}}\right]/Z\,.\end{equation}
Here $Z$ is the partition function and $\tau=1/k_{B}T$ is the inverse
temperature. In this case, the mean occupation numbers are diagonal,
and are well-known. They are given by the Fermi-Dirac distribution;\begin{equation}
\left\langle \widehat{b}_{i}^{\dagger}\widehat{b}_{j}\right\rangle =\bar{n}_{ij}=\frac{\delta_{ij}}{1+e^{\tau(\omega_{i}-\mu)}}\,.\end{equation}
However, both Gaussian forms are equivalent. A normally ordered thermal
Gaussian can always be chosen so that $\mathbf{n}$ is Hermitian,
and hence $\widehat{\rho}(\tau)=\widehat{\Lambda}(\overrightarrow{\lambda})$
if and only if $\Omega=1$ and $n_{ij}=\bar{n}_{ij}$ . 

A rather trivial example is the vacuum state, in which $\mathbf{n}=\mathbf{0}$,
so that: \begin{eqnarray}
\widehat{\Lambda}(1,\mathbf{0},\mathbf{0},\mathbf{0}) & = & \left|\mathbf{0}\right\rangle \left\langle \mathbf{0}\right|\nonumber \\
 & = & :\exp\left[-\widehat{\mathbf{b}}^{\dagger}\widehat{\mathbf{b}}\right]:\,\,.\label{Vacuum}\end{eqnarray}

We emphasize that since the Gaussian forms used here are not necessarily
Hermitian, the generalized thermal operators are a much larger set
of operators than the usual canonical thermal density matrices.

\subsection{Generalised BCS states}

A second important subset of the Gaussian operators is the generalisation
of the Bardeen-Cooper-Schreiffer (BCS) states, which are an excellent
approximation to the ground state of a weakly interacting (BCS) superconductor.
The BCS states are the fermionic equivalent of the squeezed states
found in quantum optics, and are composed only of correlated fermion
pairs. In the case of fermions, these are the fundamental pure states
that carry phase information. In Bose gases, coherent states can also
carry phase information (as in a laser or Bose-Einstein condensate),
but the fermionic equivalent of these is an unphysical Grassmann coherent
state. 

An unnormalized pure BCS state is defined as\cite{Schrieffer}:\begin{equation}
\left|\Psi_{\mathrm{BCS}}\right\rangle =\exp\left[\widehat{\mathbf{b}}^{\dagger}{\bf \mathbf{g}}\widehat{\mathbf{b}}^{\dagger}/2\right]\left|\mathbf{0}\right\rangle \,,\end{equation}
so that the corresponding density matrix is:\begin{eqnarray}
\widehat{\rho}_{\mathrm{BCS}} & = & \left|\Psi_{\mathrm{BCS}}\right\rangle \left\langle \Psi_{\mathrm{BCS}}\right|\nonumber \\
 & = & \exp\left[\widehat{\mathbf{b}}^{\dagger}{\bf \mathbf{g}}\widehat{\mathbf{b}}^{\dagger}/2\right]\left|\mathbf{0}\right\rangle \left\langle \mathbf{0}\right|\exp\left[\widehat{\mathbf{b}}{\bf \mathbf{g}}^{\dagger}\widehat{\mathbf{b}}/2\right]\nonumber \\
 & = & :\exp\left[\widehat{\mathbf{b}}^{\dagger}{\bf \mathbf{g}}\widehat{\mathbf{b}}^{\dagger}/2-\widehat{\mathbf{b}}^{\dagger}\widehat{\mathbf{b}}+\widehat{\mathbf{b}}{\bf \mathbf{g}}^{\dagger}\widehat{\mathbf{b}}/2\right]:\,.\end{eqnarray}
Apart from being unnormalized, this corresponds directly to a Gaussian
in our normal form. 

More general non-Hermitian BCS-type states are obtained on replacing
${\bf \mathbf{g}}^{\dagger}$ by an independent matrix ${\bf \mathbf{g}}^{+}$.
This generalized BCS Gaussian has an extended covariance matrix of:

\begin{equation}
\underline{\underline{\sigma}}=\left[\begin{array}{cc}
\left(\mathbf{I}+\mathbf{g}\mathbf{g}^{+}\right)^{-1} & 0\\
0 & \left(\mathbf{I}+\mathbf{g}^{+}\mathbf{g}\right)^{-1}\end{array}\right]\left[\begin{array}{cc}
-\mathbf{I} & \mathbf{g}\\
\mathbf{g}^{+} & \mathbf{I}\end{array}\right]\,\end{equation}
Clearly, from this we can see that the occupation numbers and correlations
for a generalized BCS state are given by:

\begin{eqnarray}
\mathbf{n} & = & {\bf \mathbf{g}}^{+}\left(\mathbf{I}+\mathbf{g}\mathbf{g}^{+}\right)^{-1}{\bf \mathbf{g}},\nonumber \\
\tilde{\mathbf{n}} & = & \left(\mathbf{I}+\mathbf{g}^{+}\mathbf{g}\right)^{-1}\nonumber \\
\mathbf{m} & = & \left(\mathbf{I}+\mathbf{g}\mathbf{g}^{+}\right)^{-1}{\bf \mathbf{g}}\,\nonumber \\
\mathbf{m}^{+} & = & \mathbf{g}^{+}\left(\mathbf{I}+\mathbf{g}\mathbf{g}^{+}\right)^{-1},\end{eqnarray}
 which gives the expected result that $\mathbf{m}^{+}\mathbf{m}=\tilde{\mathbf{n}}\mathbf{n}.$ 

In summary, the usual BCS states have a density matrix which is Gaussian,
and has $\mathbf{g^{+}=\mathbf{g}^{\dagger}}$. These pure states
exist as a subset of a more general class of BCS-like Gaussian operators.
This class also includes operators which have $\mathbf{g^{+}\neq\mathbf{g}^{\dagger}}$,
and are therefore not Hermitian. While these operators do not correspond
to any physical state, a linear combination of them - provided the
result is Hermitian and positive-definite - can still correspond to
a possible physical fermionic many-body state.

\section{Gaussian representation}

\label{sec:Gaussian-Representation}

While the Gaussian operators include a large and interesting set of
physical density operators, there are many cases where the existence
of interparticle interactions leads to more general fermionic states
whose correlations are of more complex, non-Gaussian forms. In all
such cases, the overall physical density operator can still be expressed
as a positive distribution over the Gaussian operators. Furthermore,
any two-body operator acting on a generalised Gaussian can be written
as a second-order derivative. These important results, proved in \cite{CorneyDrummond2005A},
means that probabilistic, random sampling methods may be used to calculate
physical observables, as we show below.

\subsection{Definition}

The Gaussian representation for fermion operators is defined as an
expansion of the density matrix for any physical state $\widehat{\rho}(t)$
as a distribution over the Gaussian basis. That is:\begin{equation}
\widehat{\rho}(t)=\int P(\overrightarrow{\lambda},t)\widehat{\Lambda}(\overrightarrow{\lambda})d\overrightarrow{\lambda}\,\,,\label{eq:general-expansion}\end{equation}
where the expansion coefficients are normalised to one: \begin{equation}
\int P(\overrightarrow{\lambda},t)d\overrightarrow{\lambda}=1\,\,.\end{equation}
This expansion defines a type of phase-space representation of the
state: the vector $\overrightarrow{\lambda}$ of Gaussian parameters
becomes a generalised phase-space coordinate, the function $P(\overrightarrow{\lambda},t)$
is then a probability distribution function over the generalised phase
space, and $d\overrightarrow{\lambda}=d^{2(p+1)}\overrightarrow{\lambda}$
is the phase-space integration measure.

\subsection{Moments}

Some basic properties of $P(\overrightarrow{\lambda},t)$ follow from
those of the Gaussian operators. For example, using the normalisation
of the Gaussian operators we find that \begin{eqnarray}
{\rm Tr}\left[\widehat{\rho}\right] & = & \int P(\overrightarrow{\lambda},t)\Omega d\overrightarrow{\lambda}\equiv\overline{\Omega}.\end{eqnarray}
Thus the normalised distribution $P$ can represent unnormalised density
operators by incorporating the normalisation into the mean weight
$\overline{\Omega}$. 

More generally, the expectation value of an operator $\widehat{O}$
evaluates to\begin{eqnarray}
\left\langle \widehat{O}\right\rangle  & \equiv & {\rm Tr}\left[\widehat{O}\,\widehat{\rho}\right]{\rm /Tr}\left[\widehat{\rho}\right]\nonumber \\
 & = & \int P(\overrightarrow{\lambda},t){\rm Tr}\left[\widehat{O}\,\widehat{\Lambda}\right]d\overrightarrow{\lambda}/\overline{\Omega}\nonumber \\
 & \equiv & \left\langle O(\overrightarrow{\lambda})\right\rangle _{P}\,,\end{eqnarray}
where the weighted average $\left\langle \ldots\right\rangle _{P}$
is defined as:\begin{equation}
\left\langle O(\overrightarrow{\lambda})\right\rangle _{P}=\int P(\overrightarrow{\lambda},t)\Omega O(\overrightarrow{\lambda})d\overrightarrow{\lambda}/\overline{\Omega}\,\,.\end{equation}
 The phase-space function $O(\overrightarrow{\lambda})$ corresponding
to the operator $\widehat{O}$ is defined as previously, using the
generalised Wick result of Eq.~(\ref{eq:Wick}). 

Physical quantities thus correspond to (weighted) moments of $P$.
For example, from traces evaluated in Sec.~\ref{sub:Moments}, we
find that the normal and anomalous Green's functions correspond to
first order moments:

\begin{eqnarray}
\left\langle \widehat{b}_{i}\widehat{b}_{j}\right\rangle  & = & \left\langle m_{ij}\right\rangle _{P}\,,\nonumber \\
\left\langle \widehat{b}_{i}^{\dagger}\widehat{b}_{j}\right\rangle  & = & \left\langle n_{ij}\right\rangle _{P}\,,\nonumber \\
\left\langle \widehat{b}_{i}^{\dagger}\widehat{b}_{j}^{\dagger}\right\rangle  & = & \left\langle m_{ij}^{+}\right\rangle _{P}\,.\label{moments}\end{eqnarray}
 Number-number correlations correspond to averages of products of
these moments:\begin{eqnarray}
\left\langle :\widehat{n}_{i}\widehat{n}_{j}:\right\rangle  & = & \left\langle n_{ii}n_{jj}-n_{ij}n_{ji}+m_{ij}^{+}m_{ji}\right\rangle _{P}\,,\end{eqnarray}
where $\widehat{n}_{i}\equiv\widehat{b}_{i}^{\dagger}\widehat{b}_{i}\,$.

Similarly, higher-order correlations correspond to higher-order moments,
the form of which are also determined by the generalised Wick result
of Eq. (\ref{eq:Wick}).

We note that the expectation value of any odd product of operators
must vanish e.g. $\left\langle \widehat{b}_{i}\right\rangle =0$.
Thus the distribution cannot represent a superposition of states whose
total number differ by an odd number. Such superposition states we
exclude from our definition of physical state, as they are not generated
by evolution under any known physical Hamiltonian. The Gaussian distribution
can, however, represent systems in which particles are coherently
added or removed in pairs, leading to nonzero anomalous correlations
$\left\langle m_{ij}\right\rangle _{P}$. On the other hand, if the
total number of particles is conserved or changed only via contact
with a thermal reservoir, then the anomalous correlations will be
identically zero and we can represent the system via an expansion
in only the thermal subset of Gaussian operators.

\section{Time evolution}

\label{sub:Time-evolution}

Here we show how these positive representations of density matrices
can be put to use. By use of these representations, any quantum evolution
arising from one and two-body interactions can be sampled by classical
stochastic processes. To see this, note that the time evolution of
a density operator is determined by a master equation, of the general
form\begin{eqnarray}
\frac{d}{dt}\widehat{\rho}(t) & = & \widehat{L}\left[\widehat{\rho}(t)\right],\label{eq:general_master_equation}\end{eqnarray}
where the $\widehat{L}$ is a superoperator that pre- and post-multiplies
the density operator by combinations of annihilation and creation
operators.

\subsection{Types of evolution}

We consider three general time-evolution categories:

\subsubsection*{Hamiltonian quantum dynamics }

For unitary evolution in real time, the superoperator is a commutator
with the Hamiltonian: \begin{eqnarray}
\widehat{L}\left[\widehat{\rho}\right] & = & -\frac{i}{\hbar}\left[\widehat{H},\widehat{\rho}\right]\,.\label{eq:dynamical_master_equation}\end{eqnarray}

\subsubsection*{Irreversible quantum dynamics}

More generally, for an open quantum system, there will be additional
terms of Lindblad form\cite{Lindblad,gardiner03} to describe the
coupling to the environment:\begin{eqnarray}
\widehat{L}\left[\widehat{\rho}\right] & = & -\frac{i}{\hbar}\left[\widehat{H},\widehat{\rho}\right]+\sum_{K}\left(2\widehat{O}_{K}\widehat{\rho}\,\widehat{O}_{K}^{\dagger}-[\widehat{O}_{K}^{\dagger}\widehat{O}_{K},\widehat{\rho}]_{+}\right),\nonumber \\
\label{eq:open_master_equation}\end{eqnarray}
where the operators $\widehat{O}_{K}$ depend on the correlations
of the environment or reservoir, within the Markov approximation.

\subsubsection*{Thermal equilibrium ensemble}

To calculate the canonical thermal equilibrium state at temperature
$T=1/k_{B}\tau$, one can solve an inverse temperature equation for
the unnormalised density operator: \begin{eqnarray}
\frac{d}{d\tau}\widehat{\rho} & = & -\frac{1}{2}[\widehat{H}-\mu\widehat{N}\,,\widehat{\rho}]_{+}\,,\label{eq:equilibrium_master_equation}\end{eqnarray}
 the solution of which will generate the unnormalised density operator
for a grand canonical distribution: $\rho(\tau)=\exp[-\tau(\widehat{H}-\mu\widehat{N})]$.

\subsection{Operator Mappings}

We wish to show how to transform a general operator time-evolution
equation (Eq.~(\ref{eq:general_master_equation})) into a Fokker-Planck
equation for the distribution, and hence into a stochastic equation.
A crucial part of this procedure is to be able to transform the operator
equations into a differential form. 

The first step is to substitute for $\widehat{\rho}$ the expansion
in Eq.~(\ref{eq:general-expansion}):\begin{eqnarray}
\int\frac{dP(\overrightarrow{\lambda},t)}{dt}\widehat{\Lambda}(\overrightarrow{\lambda})d\overrightarrow{\lambda} & = & \int P(\overrightarrow{\lambda},t)\,\widehat{L}\left[\widehat{\Lambda}(\overrightarrow{\lambda})\right]d\overrightarrow{\lambda}\,.\nonumber \\
\label{eq:expanded_master_equation}\end{eqnarray}
Second, we use the differential identities derived in \cite{CorneyDrummond2005A}
to convert the superoperator $\widehat{L}\left[\widehat{\Lambda}\right]$
into an operator $\mathcal{L}\left[\widehat{\Lambda}\right]$ that
contains only derivatives of $\widehat{\Lambda}$. Next we integrate
by parts to obtain, provided that no boundary terms arise,\begin{eqnarray}
\int\frac{dP(\overrightarrow{\lambda},t)}{dt}\widehat{\Lambda}(\overrightarrow{\lambda})d\overrightarrow{\lambda} & = & \int\mathcal{L}'\left[P(\overrightarrow{\lambda},t)\right]\,\widehat{\Lambda}(\overrightarrow{\lambda})d\overrightarrow{\lambda}\,,\nonumber \\
\label{eq:partial_integration}\end{eqnarray}
where $\mathcal{L}'$ is a reordered form of $\mathcal{L}$, with
a sign change to derivatives of odd order. Finally, we see that this
equation holds if the distribution function satisfies the evolution
equation \begin{eqnarray}
\frac{d}{dt} & P(\overrightarrow{\lambda},t)= & \mathcal{L}'\left[P(\overrightarrow{\lambda},t)\right].\label{eq:P_evolution}\end{eqnarray}

This procedure for going from the master equation for $\widehat{\rho}$
to the evolution equation for $P$ can be implemented using a set
of operator mappings, in which we introduce \emph{anti}normal ordering
as the opposite of normal ordering, and denote it via curly braces:
$\{\widehat{b}_{j}^{\dagger}\widehat{b}_{i}\}=-\widehat{b}_{i}\widehat{b}_{j}^{\dagger}\,\,$.
More generally, we can define nested orderings, in which the outer
ordering does not reorder the inner one. For example, $\{:\widehat{\rho}\widehat{b}_{j}^{\dagger}:\widehat{b}_{i}\}=-\widehat{b}_{i}\widehat{b}_{j}^{\dagger}:\widehat{\rho}:\,\,$,
where $\widehat{\rho}$ is some density operator. When ordering products
that contain the density operator $\widehat{\rho}$, we do not change
the ordering of $\widehat{\rho}$ itself; the other operators are
merely reordered around it. 

Including all possible orderings, we obtain the following mappings:

\begin{eqnarray}
\widehat{\rho} & \longrightarrow & -\frac{\partial}{\partial\Omega}\Omega P\,,\nonumber \\
:\widehat{\rho}\,\widehat{\underline{b}}\,\widehat{\underline{b}}^{\dagger}: & \longrightarrow & \left[\underline{\underline{\sigma}}^{(s)}+2\underline{\sigma}\overleftrightarrow{\frac{\partial}{\partial\underline{\underline{\sigma}}}}\,\underline{\underline{\sigma}}\right]P\,,\nonumber \\
:\left\{ \widehat{\rho}\,\widehat{\underline{b}}\right\} \widehat{\underline{b}}^{\dagger}: & \longrightarrow & \left[\widetilde{\underline{\underline{\sigma}}}^{(s)}+2\widetilde{\underline{\underline{\sigma}}}\overleftrightarrow{\frac{\partial}{\partial\underline{\underline{\sigma}}}}\,\underline{\underline{\sigma}}\right]P\,,\nonumber \\
:\widehat{\underline{b}}\left\{ \widehat{\underline{b}}^{\dagger}\widehat{\rho}\right\} : & \longrightarrow & \left[\underline{\underline{\widetilde{\sigma}}}^{(s)}+2\underline{\underline{\sigma}}\overleftrightarrow{\frac{\partial}{\partial\underline{\underline{\sigma}}}}\,\widetilde{\underline{\underline{\sigma}}}\right]P\,,\nonumber \\
\left\{ \widehat{\rho}\,\widehat{\underline{b}}\,\widehat{\underline{b}}^{\dagger}\right\}  & \longrightarrow & \left[-\widetilde{\underline{\underline{\sigma}}}^{(s)}+2\widetilde{\underline{\underline{\sigma}}}\overleftrightarrow{\frac{\partial}{\partial\underline{\underline{\sigma}}}}\,\widetilde{\underline{\underline{\sigma}}}\right]P\,.\nonumber \\
\label{eq:matrix_correspondances}\end{eqnarray}
Here, $\widetilde{\underline{\underline{\sigma}}}=\underline{\underline{I}}-\underline{\underline{\sigma}}$,
and $\underline{\underline{\sigma}}^{(s)}=\frac{1}{2}\left(\underline{\underline{\sigma}}-\underline{\underline{\sigma}}^{+}\right)$.
The notation $\overleftrightarrow{\frac{\partial}{\partial x}}$ indicates
a differentiation on both left and right sides with the ordering of
matrix multiplication preserved, so that: \begin{equation}
\left[\underline{\underline{\sigma}}\overleftrightarrow{\frac{\partial}{\partial\underline{\underline{\sigma}}}}\,\underline{\underline{\sigma}}\right]_{\mu\nu}\equiv\frac{\partial}{\partial\sigma_{\mu'\nu'}}\sigma_{\mu\nu'}\sigma_{\mu'\nu}\end{equation}

For convenience of the reader, these identities are summarized in
a more explicit form using the $M\times M$ submatrices, in the Appendix.
We note here that the mixed identities involving nested orderings
are not independent - one can always be obtained from the other. Also,
since the kernel is analytic, the distinct analytic derivatives of
the kernel are all interchangeable and lead to equivalent identities,
so that generically if $\lambda_{a}=\lambda_{a}^{x}+i\lambda_{a}^{y}$,
then $\partial/\partial\lambda_{a}=\partial/\partial\lambda_{a}^{x}=-i\partial/\partial\lambda_{a}^{y}$.
Another freedom is that $\underline{\underline{\sigma}}$ can be replaced
by $-\underline{\underline{\sigma}}^{+}$ in any of the identities.

If there are higher than quadratic terms present, the differential
mappings are applied in sequence. The operator set closest to the
operator $\widehat{\rho}\,$ leads to the innermost differential operator
acting on $P$. Thus, for example, 

\begin{eqnarray}
:\widehat{\rho}\,\widehat{b}_{\mu}\widehat{b}_{\nu}^{\dagger}\widehat{b}_{\mu'}\widehat{b}_{\nu'}^{\dagger}: & \longrightarrow & \left[\sigma_{\mu'\nu'}^{(s)}+2\frac{\partial}{\partial\sigma_{\alpha\beta}}\sigma_{\mu'\beta}\sigma_{\alpha\nu'}\right]\times\nonumber \\
 & \times & \left[\sigma_{\mu\nu}^{(s)}+2\frac{\partial}{\partial\sigma_{\gamma\delta}}\sigma_{\mu\delta}\sigma_{\gamma\nu}\right]P\,\end{eqnarray}

For a system in which the total number is conserved, one can use the
simpler thermal subset of these correspondences, i.e. including only
those that contain terms that remain when all anomalous correlations
vanish:

\begin{eqnarray}
\widehat{b}_{i}^{\dagger}\widehat{\rho}\widehat{b}_{j} & \longrightarrow & \left[\tilde{n}_{ij}-\frac{\partial}{\partial n_{lk}}\tilde{n}_{ik}\tilde{n}_{lj}\right]P\,,\nonumber \\
\widehat{b}_{i}^{\dagger}\widehat{b}_{j}\widehat{\rho} & \longrightarrow & \left[n_{ij}-\frac{\partial}{\partial n_{lk}}\tilde{n}_{ik}n_{lj}\right]P\,,\nonumber \\
\widehat{\rho}\widehat{b}_{i}^{\dagger}\widehat{b}_{j} & \longrightarrow & \left[n_{ij}-\frac{\partial}{\partial n_{lk}}n_{ik}\tilde{n}_{lj}\right]P\,,\nonumber \\
\widehat{b}_{j}\widehat{\rho}\widehat{b}_{i}^{\dagger} & \longrightarrow & \left[n_{ij}+\frac{\partial}{\partial n_{lk}}n_{ik}n_{lj}\right]P\,.\label{eq:thermal_correspondances}\end{eqnarray}

\subsection{Fokker-Planck equation}

To be able to sample the time evolution of $P$ with stochastic phase-space
equations, which is the final goal, we must have an evolution equation
that is in the form of a Fokker-Planck equation, containing first
and second order derivatives:\begin{eqnarray}
\frac{d}{dt}P(\overrightarrow{\lambda},t) & = & \left[-\sum_{a=0}^{p}\frac{\partial}{\partial\lambda_{a}}A_{a}(\overrightarrow{\lambda})\right.\label{eq:Fokker-Planck}\\
 &  & \left.+\frac{1}{2}\sum_{a,b=0}^{p}\frac{\partial}{\partial\lambda_{a}}\frac{\partial}{\partial\lambda_{b}}D_{ab}(\overrightarrow{\lambda})\right]P(\overrightarrow{\lambda},t)\,\,,\nonumber \end{eqnarray}
 where $a=0,\ldots p$ is an index that ranges over all the \emph{}variables
in the phase space. The matrix $D_{ab}$ must be positive-definite
when the Fokker-Planck equation is written in terms of real variables.
Fortunately, the fact that the representation kernel $\widehat{\Lambda}(\overrightarrow{\lambda})$
is analytic in the phase-space variables $\overrightarrow{\lambda}$
means that the matrix $D_{ab}$ can \emph{always} be chosen positive-definite
after it is divided into real and imaginary parts\cite{positiveP1},
through appropriate choices of the equivalent analytic forms $\partial/\partial\lambda_{a}=\partial/\partial\lambda_{a}^{x}=-i\partial/\partial\lambda_{a}^{y}$.

A Monte-Carlo type sampling of Eq.~(\ref{eq:Fokker-Planck}) can
be realised by integrating the Ito stochastic equations\begin{eqnarray}
d\lambda_{a}(t) & = & A_{a}(\overrightarrow{\lambda})\, dt+\sum_{b}B_{ab}(\overrightarrow{\lambda})\, dW_{b}(t)\,,\label{eq:Ito}\end{eqnarray}
where $dW_{b}(t)$ are Weiner increments, obeying $\left\langle dW_{b}(t)\, dW_{b'}(t')\right\rangle =\delta_{b,b'}\delta(t-t')dt$,
i.~e.~ Gaussian white noise. The noise matrix $B_{ab}$ is related
to the diffusion matrix by $D_{ab}=\sum_{c}B_{ac}B_{bc}\,$. This
equation is directly equivalent to a path-integral in phase-space,
so that the procedures outlined here can be regarded as a route to
obtaining a path-integral without Grassmann variables. 

Auxiliary field methods\cite{RomboutsHeyde03} can also be used to
obtain a non-Grassmann path integral, but these are generally much
more restrictive.

\subsection{Stochastic gauges}

The final phase-space equations are far from being unique. This freedom
in the final form arises from different choices that are made at different
points in the procedure. The choices at some points are constrained
by the need to generate a genuine Fokker-Planck equation with a positive-definite
diffusion matrix and vanishing boundary terms. Other than this, the
choices are in principle free; they affect the final stochastic behaviour
without changing observable moments. They are thus a stochastic analogue
of a gauge choice in field theories, and a good choice of stochastic
gauge can dramatically improve the performance of the simulations\cite{DeuarDrummond02}. 

Because the Gaussian basis is analytic, methods previously used for
the (bosonic) stochastic gauge positive-P representation are therefore
applicable\cite{DeuarDrummond02,DiffusionGauge,DeuarDrummond01}.
In the fermionic case there are three sources of gauge freedom:

\subsubsection{Fermi gauges}

For fermionic systems there is a freedom in the choice of operator
correspondences, arising from vanishing operator products; any term
involving a square of a fermion operator, like $\widehat{a}_{i}^{2}\widehat{O}$,
is zero. Terms like this (and products of such terms), can be added
to the Hamiltonian or Liouville equation without modifying the density
matrix. The corresponding additional differential terms may not vanish,
hence generating a different but equivalent stochastic equation. Such
a fermionic stochastic gauge is necessary to avoid complex weights
in imaginary-time simulations of interacting systems, such as the
Hubbard model\cite{GaussianQMC}.

\subsubsection{Diffusion gauges}

Diffusions gauges arise from the fact that the matrix square root
$D_{ab}=\sum_{c}B_{ac}B_{bc}\,$has multiple solutions, especially
if one notes that there is no restriction on the second dimension
of $B_{ab}$. This changes the stochastic noise term and can lead
to a reduction in sampling error\cite{DiffusionGauge}.

\subsubsection{Drift gauge}

Drift gauges are obtained by trading off trajectory weight against
trajectory direction. The possibility for drift gauges arises from
the weight $\Omega$ in the density-operator expansion. The first
of the correspondences in Eq.~(\ref{eq:matrix_correspondances})
can be used to convert drift terms for the phase-space variables into
diffusion terms for the weight\cite{ICOLSProceedings}. As a result,
one can add an arbitrary gauge $g_{a}(\overrightarrow{\lambda})$,
of the same dimension as the noise vector. Assuming $B_{0b}=0$, and
using Einstein summation conventions, this leads to:

\begin{eqnarray}
d\Omega(t) & = & A_{0}dt+\Omega g_{b}dW_{b}(t)\,,\label{eq:Gauge}\\
d\lambda_{a}(t) & = & A_{a}\, dt+B_{ab}\left[\, dW_{b}(t)-g_{b}dt\,\right].\nonumber \end{eqnarray}
Previous work\cite{DeuarDrummond01,DeuarDrummond02} has shown that
drift gauges can remove boundary terms in bosonic positive-$P$ representation
by stabilizing deterministic trajectories.

\section{Examples}

\label{sec:Examples}

The virtue of phase-space representation is that while Hilbert space
dimension grows exponentially with the number of modes $M$, the phase-space
dimension only grows quadratically. Thus, for example, a problem involving
$M=1000$ fermion modes has a Hilbert space dimension of $D=2^{1000}=10^{103}$
dimensions. This is larger than the number of particles in the observable
universe (which is perhaps $10^{85}$ by current astrophysical reckoning).
By contrast, the fermion phase-space dimension is $4\times10^{6}$.
While large, this is not astronomical.

Hamiltonians and general time-evolution equations that are only quadratic
in the Fermi ladder operators, i.~e.~constructed from one-body operators,
will map to a Fokker-Planck equation that contains only first order
derivatives. The evolving quantum state can thus be sampled by a single,
deterministic trajectory. More generally, quartic terms and cubic
terms (if Bose operators are included) can also be handled, and these
result in stochastic equations or their equivalent path integrals.

Examples of how some typical Fermi problems are mapped into phase-space
equations are given as follows.

\subsection{Free gas}

As an example of quadratic evolution, consider the thermal equilibrium
calculation for a gas of noninteracting particles. The governing Hamiltonian
(including the chemical potential) is always diagonalizable, and can
be written as:

\begin{equation}
\widehat{H}=\widehat{\mathbf{b}}^{\dagger}\bm\omega\,\widehat{\mathbf{b}}\,,\label{eq:thermalHamiltonian}\end{equation}
where $\omega_{ij}=\delta_{ij}\omega_{j}$are the single-particle
energies. The grand canonical distribution at temperature $T=1/k_{B}\tau$
is found from the equation\begin{eqnarray}
\frac{\partial}{\partial\tau}\widehat{\rho} & = & -\frac{1}{2}\left(\widehat{\mathbf{b}}^{\dagger}\bm\omega\,\widehat{\mathbf{b}}\,\widehat{\rho}+\widehat{\rho}\,\widehat{\mathbf{b}}^{\dagger}\bm\omega\,\widehat{\mathbf{b}}\right)\,.\end{eqnarray}
 Now this master equation can be mapped to an equivalent equation
for the the distribution $P$ by use of the thermal correspondences
in Eq.~(\ref{eq:mn_correspondances}). However, because the solution
is an unnormalised density operator, there will be zeroth-order terms
in the equation. We can convert such terms to first order by applying
the weight ($\Omega)$ identity in Eq.~(\ref{eq:matrix_correspondances}),
thus obtaining the Fokker-Planck equation\begin{eqnarray}
\frac{\partial P}{\partial\tau} & = & \sum_{k}\omega_{k}\left[\frac{\partial}{\partial n_{k}}(1-n_{k})+\frac{\partial}{\partial\Omega}\Omega\right]n_{k}P\,\,.\end{eqnarray}
This Fokker-Planck equation with first-order derivatives corresponds
to deterministic characteristic equations:

\begin{eqnarray}
\dot{\Omega} & = & -\sum_{k}\omega_{k}\Omega n_{k}\,\,,\\
\dot{n}_{k} & = & -\omega_{k}n_{k}\left(1-n_{k}\right)\,\,.\end{eqnarray}

Integrating the deterministic equation for the mode occupation $n_{k}$
leads to the usual Fermi-Dirac distribution:

\begin{equation}
n_{k}=\frac{1}{e^{\omega_{k}\tau}+1}\,\,.\end{equation}
From integration of the weight equation, one finds that that normalisation
of the density operator is

\begin{equation}
\mathrm{Tr}\left[\widehat{\rho}_{u}\right]=\Omega(\tau)=\Omega_{0}\Pi_{k}e^{-\omega_{k}n_{k}\tau}\,\,,\end{equation}
 i.~e.~ the weight decays exponentially, at a rate given by the
total energy.

\subsection{General quadratic evolution }

More generally, one can have a quadratic Liouville operator in situations
involving non-thermal terms like $\hat{b}_{i}\hat{b}_{j}$. This can
occur for, example, when fermion pairs are generated from molecule
or exciton dissociation. These are even associated with certain spin-chain
problems\cite{SpinChain}, where the Jordan-Wigner theorem is used
to transform spins to fermion operators. Other quadratic Liouville
operators are commonly found in cases involving coupling to reservoirs\cite{gardiner03}. 

The generic phase-space equations for a general Fermi system with
a quadratic Liouville operator can be easily obtained, for evolution
both through time and through inverse temperature. The most general
master equation that covers both kinds of evolution can be written\begin{eqnarray}
\frac{d}{d\tau}\widehat{\rho} & = & K\widehat{\rho}-\frac{1}{2}\sum_{\mu\nu}\left({\cal A}{}_{\nu\mu}:\widehat{b}_{\mu}\widehat{b}_{\nu}^{\dagger}\widehat{\rho}:+{\cal B}{}_{\nu\mu}\left\{ \widehat{b}_{\mu}\widehat{b}_{\nu}^{\dagger}\widehat{\rho}\right\} +\right.\nonumber \\
 &  & \left.{\cal C}_{\nu\mu}:\left\{ \widehat{\rho}\widehat{b}_{\mu}\right\} \widehat{b}_{\nu}^{\dagger}:+{\cal C}_{\mu\nu}^{*}\left\{ :\widehat{\rho}\widehat{b}_{\mu}:\widehat{b}_{\nu}^{\dagger}\right\} \right),\label{eq:general_quadratic_me}\end{eqnarray}
 where the elements of $2M\times2M$ matrices $\underline{\underline{{\cal A}}}$,
$\underline{\underline{{\cal B}}}$ and $\underline{\underline{{\cal C}}}$
are determined by the coefficients of the Hamiltonian or master equation.
By applying the mappings of Eq.~(\ref{eq:matrix_correspondances}),
we find the evolution of the covariance matrix to be:\begin{eqnarray}
\frac{d}{d\tau}\underline{\underline{\sigma}} & = & \underline{\underline{\sigma}}\,\underline{\underline{{\cal A}}}\,\underline{\underline{\sigma}}+\widetilde{\underline{\underline{\sigma}}}\,\underline{\underline{{\cal B}}}\,\widetilde{\underline{\underline{\sigma}}}+\underline{\underline{\sigma}}\,\underline{\underline{{\cal C}}}\,\widetilde{\underline{\underline{\sigma}}}+\widetilde{\underline{\underline{\sigma}}}\,\underline{\underline{{\cal C}}}^{\dagger}\,\underline{\underline{\sigma}},\nonumber \\
\label{eq:general_quadratic_eq}\end{eqnarray}
 This equation simply corresponds to the characteristic or drift equations
given by the vector $\vec{A}$ in the Ito stochastic equation (\ref{eq:Ito}),
and in these cases there is no diffusion or stochastic term. Unlike
a conventional path integral, we see that a quadratic Hamiltonian
or Liouville equation simply results in a noise-free, deterministic
trajectory on phase space. For deterministic evolution such as this,
the weight $\Omega$ does not affect physical observables, so we do
not consider it here.

In the examples that follow, we assume for simplicity (but without
loss of generality) that the constant matrices have been chosen with
hermitian anti-symmetry such that: \begin{eqnarray}
\underline{\underline{A}} & = & -\underline{\underline{A}}^{+}\nonumber \\
\underline{\underline{B}} & = & -\underline{\underline{B}}^{+}\nonumber \\
\underline{\underline{C}}^{\dagger} & = & -\underline{\underline{C}}^{+}\,\,.\end{eqnarray}

\subsubsection{Temperature evolution}

For temperature evolution, the structure of the master equation (Eq.~(\ref{eq:equilibrium_master_equation}))
is such that $\underline{\underline{A}}=\underline{\underline{B}}$
and $\underline{\underline{{\cal C}}}=\underline{\underline{{\cal C}}}{}^{\dagger}$,
giving the simpler result:\begin{eqnarray}
\frac{d}{d\tau}\underline{\underline{\sigma}} & = & \frac{1}{2}\left(\underline{\underline{I}}-2\underline{\underline{\sigma}}\right)\underline{\underline{T}}\left(\underline{\underline{I}}-2\underline{\underline{\sigma}}\right)+\underline{\underline{\sigma}}^{0},\label{eq:det_equil}\end{eqnarray}
where we have introduced:\begin{eqnarray}
\underline{\underline{T}} & = & \underline{\underline{{\cal B}}}-\underline{\underline{{\cal C}}}\nonumber \\
\underline{\underline{\sigma}}^{0} & = & \frac{1}{4}\underline{\underline{I}}\left(\underline{\underline{{\cal B}}}-2\underline{\underline{{\cal C}}}\right)\underline{\underline{I}}\,\,.\end{eqnarray}

For the case of a number conserving Hamiltonian $H=\mathbf{b}^{\dagger}\bm\omega\mathbf{b}$,
we find that $\underline{\underline{B}}=0$ and \begin{eqnarray}
\underline{\underline{{\cal C}}} & = & \frac{1}{2}\left[\begin{array}{cc}
-\bm\omega^{T} & \mathbf{0}\\
\mathbf{0} & \bm\omega\end{array}\right].\end{eqnarray}
The phase-space equations then reduce to \begin{eqnarray}
\frac{d}{d\tau}\mathbf{n} & = & -\frac{1}{2}\left(\mathbf{n}\bm\omega\widetilde{\mathbf{n}}+\widetilde{\mathbf{n}}\bm\omega\mathbf{n}\right),\end{eqnarray}
which reproduces the free gas example above.

\subsubsection{Dynamical evolution}

For time evolution, with possible coupling to the environment, there
is a different symmetry to the master equation (Eq.~(\ref{eq:open_master_equation}))
that means that $\underline{\underline{{\cal A}}}+\underline{\underline{{\cal B}}}-\underline{\underline{{\cal C}}}-\underline{\underline{{\cal C}}}{}^{\dagger}=\underline{\underline{0}}$.
A formal solution to the phase-space equations can now be explicitly
written down:\begin{eqnarray}
\underline{\underline{\sigma}}(t) & = & \exp\left(-\underline{\underline{U}}^{\dagger}t\right)\left(\underline{\underline{\sigma}}(0)-\underline{\underline{\sigma}}^{\infty}\right)\exp\left(-\underline{\underline{U}}t\right)+\underline{\underline{\sigma}}^{\infty},\nonumber \\
\label{eq:det_dynamics}\end{eqnarray}
where $\underline{\underline{U}}=\left(\underline{\underline{{\cal B}}}-\underline{\underline{{\cal C}}}\right)\underline{\underline{I}}$
and where $\underline{\underline{\sigma}}^{\infty}$ satisfies: \begin{equation}
\underline{\underline{I}}\,\underline{\underline{{\cal B}}}\,\underline{\underline{I}}=\underline{\underline{U}}^{\dagger}\,\underline{\underline{\sigma}}^{\infty}+\underline{\underline{\sigma}}^{\infty}\underline{\underline{U}}\,\,.\end{equation}

To illustrate the physical meaning of these matrices, we consider
the simplest model of a small quantum dot coupled to a zero-temperature
reservoir:\begin{eqnarray}
\dot{\widehat{\rho}} & = & -i\omega\widehat{b}^{\dagger}\widehat{b}\widehat{\rho}+i\omega\widehat{\rho}\widehat{b}^{\dagger}\widehat{b}+\gamma\left(\widehat{b}\widehat{\rho}\widehat{b}^{\dagger}-\frac{1}{2}\widehat{b}^{\dagger}\widehat{b}\widehat{\rho}-\frac{1}{2}\widehat{\rho}\widehat{b}^{\dagger}\widehat{b}\right).\nonumber \\
\end{eqnarray}
 In terms of the general form, this corresponds to $\underline{\underline{{\cal A}}}=\underline{\underline{0}}$,
$\underline{\underline{{\cal B}}}=\gamma\underline{\underline{I}}$,
and \begin{eqnarray}
\underline{\underline{{\cal C}}} & = & \left[\begin{array}{cc}
-i\omega-\frac{1}{2}\gamma & 0\\
0 & -i\omega+\frac{1}{2}\gamma\end{array}\right].\end{eqnarray}
 The general solution then reduces to \begin{eqnarray}
\underline{\underline{\sigma}}(t) & = & \left[\begin{array}{cc}
e^{-i\omega-\gamma/2} & 0\\
0 & e^{i\omega+\gamma/2}\end{array}\right]\left(\underline{\underline{\sigma}}(0)-\underline{\underline{I}}\right)\nonumber \\
 &  & \times\left[\begin{array}{cc}
e^{i\omega-\gamma/2} & 0\\
0 & e^{-i\omega-\gamma/2}\end{array}\right]+\underline{\underline{I}},\end{eqnarray}
which implies that the density decays as $n(t)=e^{-\gamma t}n(0)$,
as expected. 

The solution to a multimode quantum dot model also follows from Eq.~(\ref{eq:det_dynamics}).
The relevant master equation is\begin{eqnarray}
\dot{\widehat{\rho}} & = & -i\omega_{ji}\widehat{b}_{i}^{\dagger}\widehat{b}_{j}\widehat{\rho}+i\omega_{ji}\widehat{\rho}\widehat{b}_{i}^{\dagger}\widehat{b}_{j}\nonumber \\
 &  & +\gamma_{ij}\left(\widehat{b}_{i}\widehat{\rho}\widehat{b}_{j}^{\dagger}-\frac{1}{2}\widehat{b}_{j}^{\dagger}\widehat{b}_{i}\widehat{\rho}-\frac{1}{2}\widehat{\rho}\widehat{b}_{j}^{\dagger}\widehat{b}_{i}\right),\end{eqnarray}
for which the evolution matrix is\begin{eqnarray}
\underline{\underline{U}} & = & \left[\begin{array}{cc}
e^{-i\bm\omega^{T}+\bm\gamma^{T}/2} & \mathbf{0}\\
\mathbf{0} & e^{i\bm\omega+\bm\gamma/2}\end{array}\right].\end{eqnarray}
Physically, this corresponds, as expected, to damped oscillatory behavior
(taking $\bm\gamma$ to be positive definite) in the moments:\begin{eqnarray}
\mathbf{n} & = & e^{i\bm\omega-\bm\gamma/2}\mathbf{n}(0)e^{-i\bm\omega-\bm\gamma/2},\nonumber \\
\mathbf{m} & = & e^{-i\bm\omega^{T}-\bm\gamma^{T}/2}\mathbf{m}(0)e^{-i\bm\omega-\bm\gamma/2}.\end{eqnarray}
Here, of course, there are no electron-electron interactions included.
However, such interactions can be dealt with via a stochastic sampling
methods, as we show in the next section.

\subsection{Interacting gas}

\subsubsection{Two-body interactions}

For systems of particles with two-body interactions, the Gaussian
representation gives nonlinear, stochastic phase-space equations,
which must be solved numerically. Consider a two-body interaction
of the form:\begin{eqnarray*}
\widehat{H}_{2} & = & \sum_{ij}U_{ij}\widehat{n}_{i}\widehat{n}_{j}\,,\end{eqnarray*}
where $\widehat{n}_{ij}=\widehat{a}_{i}^{\dagger}\widehat{a}_{j}$.
For a number-conserving system, we can use correspondences of Eq.~(\ref{eq:thermal_correspondances})
to generate a Fokker-Planck equation for the grand canonical evolution.
The diffusion matrix $D_{u,v}$ in this equation is\begin{eqnarray}
D_{ij,kl} & = & -\sum_{pq}U_{pq}\left\{ n_{ip}\widetilde{n}_{pj}n_{kq}\widetilde{n}_{ql}\right.\nonumber \\
 &  & \left.+\widetilde{n}_{ip}n_{pj}\widetilde{n}_{kq}n_{ql}\right\} .\end{eqnarray}
 Suppose that the interaction matrix $U_{pq}$ is negative-definite,
such that we can write it as a sum of negative squares: $U_{pq}=-\sum_{\alpha}b_{p,\alpha}b_{q,\alpha}$.
Then the diffusion matrix is positive definite, as it can be written
in the form:\begin{eqnarray}
D_{ij,kl} & = & \sum_{\alpha}\left\{ B_{ij,\alpha}^{(1)}B_{kl,\alpha}^{(1)}+B_{ij,\alpha}^{(2)}B_{kl,\alpha}^{(2)}\right\} ,\label{eq:positive-definite}\end{eqnarray}
 where the noise matrices are:\begin{eqnarray}
B_{ij,\alpha}^{(1)} & = & \sum_{p}b_{p,\alpha}n_{ip}\widetilde{n}_{pj}\,,\nonumber \\
B_{ij,\alpha}^{(2)} & = & \sum_{p}b_{p,\alpha}\widetilde{n}_{ip}n_{pj}\,.\end{eqnarray}
 Thus for an interaction of this type, the noise terms in the final
stochastic equations will be real. The form of noise terms for a more
general interaction is considered in \cite{CorbozTroyer}.

\subsubsection{Hubbard model}

As an example, we apply the representation to the Hubbard model, which
is the simplest nontrivial model for strongly interacting fermions
on a lattice. It is an important system in condensed matter physics,
with relevance to the theory of high-temperature superconductors\cite{Linden92},
and in ultracold atomic physics. The full phase-diagram in two dimensions
is not fully understood as yet. Due to developments in atomic lattices,
this model is directly experimentally accessible\cite{LatticeFermiGas,Hofstetter02}. 

The Hamiltonian for the model is\cite{Hubbard63}:\begin{eqnarray}
H(\widehat{\mathbf{n}}_{1},\widehat{\mathbf{n}}_{-1}) & = & -\sum_{ij,\sigma}t_{ij}\widehat{n}_{ij,\sigma}+U\sum_{j}\widehat{n}_{jj,1}\widehat{n}_{jj,-1},\nonumber \\
\label{eq:Hubbard}\end{eqnarray}
 where $\widehat{n}_{ij,\sigma}=\widehat{a}_{i,\sigma}^{\dagger}\widehat{a}_{j,\sigma}$=$\left\{ \widehat{\mathbf{n}}_{\sigma}\right\} _{ij}$.
The index $\sigma$ denotes spin ($\pm1$), the indices $i,j$ label
lattice location. Here $t_{ij}=t$ if the $i,j$ correspond to nearest
neighbour sites, $t_{ij}=\mu$ if $i=j$ and is otherwise $0$. The
chemical potential $\mu$ is included to control the total particle
number. 

Because the Hubbard model conserves total number and spin, one can
map this problem to a reduced phase space of $\lambda=(\Omega,n_{ij,1},n_{ij,-1})$.
Thus the simpler mappings of Eq.~(\ref{eq:thermal_correspondances})
can be used for each spin component. The one-body terms generate drift
terms only, and can be dealt with as above. The two-body terms generate
both drift and diffusion terms. Applying the mappings directly to
the Hubbard model as written above, we obtain the diffusion matrix\begin{eqnarray}
D_{ij\sigma,kl\sigma'} & = & -U\delta_{\sigma,-\sigma'}\sum_{p}\left\{ n_{ip\sigma}\widetilde{n}_{pj\sigma}n_{kp\sigma'}\widetilde{n}_{pl\sigma'}\right.\nonumber \\
 &  & \left.+\widetilde{n}_{ip\sigma}n_{pj\sigma}\widetilde{n}_{kp\sigma'}n_{pl\sigma'}\right\} ,\end{eqnarray}
which, because it has zeros on the diagonal, cannot be put into a
positive definite form with real variables. 

However, using the anticommuting properties of the Fermi operators,
we can rewrite the interaction term in the Hubbard Hamiltonian as\begin{eqnarray}
H_{I} & = & -\frac{|U|}{2}\sum_{j}:\left(\widehat{n}_{jj,1}-S\widehat{n}_{jj,-1}\right)^{2}:\,\nonumber \\
 & = & \sum_{j}U_{i\sigma,j\sigma'}:\widehat{n}_{ii,\sigma}\widehat{n}_{jj,\sigma'}\end{eqnarray}
where $S=U/|U|=\pm1$. Now in this form, the interaction matrix is
negative definite: \begin{eqnarray}
U_{i\sigma,j\sigma'} & = & -\frac{|U|}{2}\delta_{ij}\left(\delta_{\sigma,\sigma'}-S\delta_{\sigma,-\sigma'}\right)\nonumber \\
 & = & -\frac{|U|}{2}\sum_{k}\delta_{i,k}\sigma^{s}\delta_{j,k}\sigma'^{s},\end{eqnarray}
 where $s=(S+1)/2$, so that $s=0$ for the attractive case and $s=1$
for the repulsive case.

From Eq.~(\ref{eq:positive-definite}) the diffusion matrix is positive
definite, with corresponding noise matrices:\begin{eqnarray}
B_{ij\sigma,\alpha}^{(1)} & = & \sqrt{\left|U\right|/2}\sigma^{s}n_{i\alpha}\widetilde{n}_{\alpha j}\,,\nonumber \\
B_{ij\sigma,\alpha}^{(2)} & = & \sqrt{\left|U\right|/2}\sigma^{s}\widetilde{n}_{i\alpha}n_{\alpha j}\,.\end{eqnarray}

With this choice of noise terms, the final phase-space equations are,
in It\^o form,\begin{eqnarray}
\frac{d\mathbf{n}_{\sigma}}{d\tau} & = & \frac{1}{2}\left\{ \tilde{\mathbf{n}}_{\sigma}\bm T_{\sigma}^{(1)}\!\mathbf{n}_{\sigma}+\mathbf{n}_{\sigma}\bm T_{\sigma}^{(2)}\!\tilde{\mathbf{n}}_{\sigma}\right\} ,\end{eqnarray}
 where we have introduced the stochastic propagation matrix:\begin{eqnarray}
T_{ij,\sigma}^{(r)} & = & t_{ij}-\delta_{ij}\left\{ Un_{jj,-\sigma}+\sigma^{s}\xi_{j}^{(r)}\right\} .\end{eqnarray}
 The real Gaussian noise $\xi_{j}^{(r)}(\tau)$ is defined by the
correlations \begin{eqnarray*}
\left\langle \xi_{j}^{(r)}(\tau)\,\xi_{j'}^{(r')}(\tau')\right\rangle  & = & 2\left|U\right|\delta(\tau-\tau')\delta_{jj'}\delta_{rr'}\,.\,\end{eqnarray*}
Because the diffusion can be realised in terms of real noise, the
phase-space equations will not be driven off the real manifold. This
has an important implication for the weight $\Omega$, which enters
the problem because the solution will be an unnormalised density operator.
The weights for each trajectory evolve as physically expected for
energy-weighted averages, with weights depending exponentially on
the inverse temperature $\tau$ and the effective trajectory Hamiltonian
$H$: \[
\frac{d\Omega}{d\tau}=-\Omega H(\mathbf{n}_{1},\mathbf{n}_{-1})\,\,.\]
Because the equations for the phase-space variables $n_{ij,\sigma}$
are all real, the weights will all remain positive, thereby eliminating
the traditional manifestation of the sign problem. 

This method can calculate any correlation function, at any temperature,
to the precision allowed by the sampling error and subject to there
being no boundary terms in Eq.~(\ref{eq:partial_integration}). Preliminary
simulations in one\cite{JModOptics}, two\cite{GaussianQMC} and three
dimensions showed that sampling error is well-controlled, even for
very low temperatures. However, more extensive simulations of the
2D Hubbard model have shown that, at half filling, certain correlation
functions do not appear to converge to the correct zero-temperature
results at these very low temperatures\cite{CorbozTroyer}. Because
the Gaussian basis does not possess many of the symmetries of the
Hubbard model, they must be restored in the distribution over Gaussian
basis elements. For finite sampling, this restoration may be incomplete,
giving the departure from exact results at low temperatures. It has
been shown that the correct results can be obtained by applying a
projection onto a symmetric subspace\cite{CorbozTroyer}. There may
also be systematic errors if boundary terms are present. Both of these
possibilities imply that further optimization via stochastic gauge
choices may be required to keep the low-temperature distributions
compact, free from tails and from features that would lead to biasing.

\subsubsection{Drift gauges}

For the Hubbard model, we can modify the drift part according to Eq.~(\ref{eq:Gauge})
by adding a term $\mathbf{G}_{\sigma}$ to the stochastic propagation
matrices $\mathbf{T}_{\sigma}^{(r)}$. Because of the diagonal nature
of the noise terms, the added term will also be diagonal: $G_{ij,\sigma}=\delta_{ij}G_{j,\sigma}$.
The additional diffusion term in the weight equation is then\begin{eqnarray}
\left(\frac{d\Omega}{d\tau}\right)_{g} & = & \frac{\Omega}{2\left|U\right|}\sum_{jr\sigma}\sigma^{s}G_{j,\sigma}\xi_{j}^{(r)}\,.\end{eqnarray}
The choice of gauge term $\mathbf{G}_{\sigma}$ is guided on the one
hand by the need to ensure the phase-space distribution remains bounded
and on the other by the requirement of introducing only the minimum
amount of diffusion into the weight. The function should thus act
only when necessary to control large trajectories and should be zero
otherwise. 

The diagonal form of gauge term possible is able to remove instabilities
in the Hubbard equations that are directly due to the interaction
term $U$. However, instabilities may still arise from the coupling
terms $t_{ij}$, even though they are of lower order. Thus it may
be necessary to introduce weaker, off-diagonal gauge terms. This in
turn requires additional, off-diagonal noise terms in the propagation
matrix. Such noises can be introduced by use of additional Fermi gauges.
For example, the vanishing term\cite{everything}\begin{eqnarray}
0 & = & \sum_{ij\sigma}\frac{1}{2}V_{ji,\sigma}\left\{ \left(\delta_{ij}-\widehat{n}_{ij,\sigma}\right)\widehat{n}_{ij,\sigma}\widehat{\rho}+\widehat{\rho}\,\widehat{n}_{ij,\sigma}\left(\delta_{ij}-\widehat{n}_{ij,\sigma}\right)\right\} ,\nonumber \\
\end{eqnarray}
 where $V_{ij,\sigma}$ are positive numbers, gives the additional
stochastic contribution to the propagation matrix:\begin{eqnarray}
T_{ij,\sigma}^{(r)} & \rightarrow & T_{ij,\sigma}^{(r)}+\zeta_{ij,\sigma}^{(r)}(\tau)\,,\end{eqnarray}
where the new noises $\zeta_{ij,\sigma}^{(r)}(\tau)$ have the correlations\begin{eqnarray}
\left\langle \zeta_{ij}^{(r)}(\tau)\,\zeta_{i'j'}^{(r')}(\tau')\right\rangle  & = & 4V_{ij,\sigma}\delta(\tau-\tau')\delta_{ii'}\delta_{jj'}\delta_{rr'}\,.\end{eqnarray}
We can now introduce arbitrary off-diagonal gauge terms $G_{ij,\sigma}$
into the propagation matrix, with the corresponding diffusion term
in the weight equation\begin{eqnarray}
\left(\frac{d\Omega}{d\tau}\right)_{g} & = & -\Omega\sum_{ijr\sigma}G_{ij,\sigma}\zeta_{ij}^{(r)}/4V_{ij,\sigma}\,.\end{eqnarray}
 Again there is a trade-off between gauge strength and additional
diffusion. But there is also a freedom (in the choice of $V_{ij,\sigma}$)
as to whether the noise appears in the weight equation or in the propagation
matrix. 

With such a combination of Fermi and drift gauges, it is possible
to introduce terms to stabilise the drift evolution of any of the
phase-space variables $n_{ij,\sigma}$, and so maintain a bounded
phase-space distribution.

\section{Conclusion}

In summary, we have introduced a phase-space representation for many-body
fermionic states, enabling new types of first-principles calculations
and simulations of highly correlated systems. Many-body systems with
one- and two-body interactions can be solved by use of stochastic
sampling methods, since they can be transformed into a second-order
Fokker-Planck equation, provided a suitable stochastic gauge is chosen
to ensure that the distribution remains sufficiently bounded. 

These techniques are potentially applicable to a wide range of fermionic
problems, including both real-time and finite temperature calculations.
Generalized master equations for non-equilibrium fermionic open systems
coupled to reservoirs are a particularly suitable type of application.
We have given examples of the use of fermionic differential identities
to transform multi-mode master equations into deterministic phase-space
equations, although more general interactions typically lead to stochastic
equations. These equations have exponentially less complexity than
the full Hilbert space equations, are generally simpler to solve than
path integrals, and never involve either Grassmann variables or determinants.

The application to the Hubbard model demonstrates the immediate utility
of the Gaussian method to solving long-standing problems in many-body
quantum physics, provided suitable gauges can be found to ensure that
boundary terms to not arise. Rapid experimental advances in the area
of ultra-cold fermionic atoms\cite{Experiments} mean that direct
and quantitative tests of precise theoretical predictions should be
feasible in the near future. Demonstration of a quantum degenerate
Fermi gas in a lattice has already taken place\cite{LatticeFermiGas}. 

The general technique established here potentially also has broad
applicability in many other areas of quantum many-body theory and
quantum field theory.

\begin{acknowledgments}
Funding for this research was provided by an Australian Research Council
Centre of Excellence grant. We also acknowledge useful discussions
with R. J. Glauber, I. Cirac, F. F. Assad, M. Troyer, T. Esslinger,
S. Rombouts and K. V. Kheruntsyan.
\end{acknowledgments}
\newpage
\appendix

\section*{Appendix}

It is sometimes more convenient to work with explicit $\mathbf{n},\mathbf{m},\mathbf{m}^{+}$submatrices
rather than the total covariance. In fully indexed notation, using
the $M\times M$ submatrices, the Fermi operator correspondences (Eq.~(\ref{eq:matrix_correspondances}))become:\onecolumngrid

\begin{eqnarray}
\widehat{b}_{i}^{\dagger}\widehat{b}_{j}\widehat{\rho} & \longrightarrow & \left[n_{ij}-\frac{\partial}{\partial n_{lk}}\left\{ n_{lj}\tilde{n}_{ik}+m_{li}^{+}m_{jk}\right\} -\frac{\partial}{\partial m_{lk}}\left\{ m_{lj}\tilde{n}_{ik}+\tilde{n}_{il}m_{jk}\right\} +\frac{\partial}{\partial m_{lk}^{+}}\left\{ n_{lj}m_{ik}^{+}+m_{li}^{+}n_{kj}\right\} \right]P\,,\nonumber \\
\widehat{\rho}\widehat{b}_{i}^{\dagger}\widehat{b}_{j} & \longrightarrow & \left[n_{ij}-\frac{\partial}{\partial n_{lk}}\left\{ \tilde{n}_{lj}n_{ik}+m_{li}^{+}m_{jk}\right\} +\frac{\partial}{\partial m_{lk}}\left\{ m_{lj}n_{ik}+n_{il}m_{jk}\right\} -\frac{\partial}{\partial m_{lk}^{+}}\left\{ \tilde{n}_{lj}m_{ik}^{+}+m_{li}^{+}\tilde{n}_{kj}\right\} \right]P\,,\nonumber \\
\widehat{b}_{i}^{\dagger}\widehat{\rho}\widehat{b}_{j} & \longrightarrow & \left[\tilde{n}_{ij}-\frac{\partial}{\partial n_{lk}}\left\{ \tilde{n}_{lj}\tilde{n}_{ik}-m_{li}^{+}m_{jk}\right\} +\frac{\partial}{\partial m_{lk}}\left\{ m_{lj}\tilde{n}_{ik}+\tilde{n}_{il}m_{jk}\right\} +\frac{\partial}{\partial m_{lk}^{+}}\left\{ \tilde{n}_{lj}m_{ik}^{+}+m_{li}^{+}\tilde{n}_{kj}\right\} \right]P\,,\nonumber \\
\widehat{b}_{j}\widehat{\rho}\widehat{b}_{i}^{\dagger} & \longrightarrow & \left[n_{ij}-\frac{\partial}{\partial n_{lk}}\left\{ m_{li}^{+}m_{jk}-n_{lj}n_{ik}\right\} +\frac{\partial}{\partial m_{lk}}\left\{ m_{lj}n_{ik}+n_{il}m_{jk}\right\} +\frac{\partial}{\partial m_{lk}^{+}}\left\{ n_{lj}m_{ik}^{+}+m_{li}^{+}n_{kj}\right\} \right]P\,,\nonumber \\
\widehat{b}_{i}\widehat{b}_{j}\widehat{\rho} & \longrightarrow & \left[m_{ij}-\frac{\partial}{\partial n_{lk}}\left\{ n_{li}m_{jk}-n_{lj}m_{ik}\right\} -\frac{\partial}{\partial m_{lk}}\left\{ m_{li}m_{jk}-m_{lj}m_{ik}\right\} -\frac{\partial}{\partial m_{lk}^{+}}\left\{ n_{lj}n_{ki}-n_{li}n_{kj}\right\} \right]P\,,\nonumber \\
\widehat{\rho}\widehat{b}_{i}\widehat{b}_{j} & \longrightarrow & \left[m_{ij}-\frac{\partial}{\partial n_{lk}}\left\{ \tilde{n}_{lj}m_{ik}-\tilde{n}_{li}m_{jk}\right\} -\frac{\partial}{\partial m_{lk}}\left\{ m_{li}m_{jk}-m_{lj}m_{ik}\right\} -\frac{\partial}{\partial m_{lk}^{+}}\left\{ \tilde{n}_{lj}\tilde{n}_{ki}-\tilde{n}_{li}\tilde{n}_{kj}\right\} \right]P\,,\nonumber \\
\widehat{b}_{j}\widehat{\rho}\widehat{b}_{i} & \longrightarrow & \left[m_{ij}+\frac{\partial}{\partial n_{lk}}\left\{ \tilde{n}_{li}m_{jk}+n_{lj}m_{ik}\right\} -\frac{\partial}{\partial m_{lk}}\left\{ m_{li}m_{jk}-m_{lj}m_{ik}\right\} -\frac{\partial}{\partial m_{lk}^{+}}\left\{ \tilde{n}_{li}n_{kj}-\tilde{n}_{lj}n_{ki}\right\} \right]P\,,\nonumber \\
\widehat{b}_{i}^{\dagger}\widehat{b}_{j}^{\dagger}\widehat{\rho} & \longrightarrow & \left[m_{ij}^{+}-\frac{\partial}{\partial n_{lk}}\left\{ m_{lj}^{+}\tilde{n}_{ik}-m_{li}^{+}\tilde{n}_{jk}\right\} -\frac{\partial}{\partial m_{lk}}\left\{ \tilde{n}_{jl}\tilde{n}_{ik}-\tilde{n}_{il}\tilde{n}_{jk}\right\} -\frac{\partial}{\partial m_{lk}^{+}}\left\{ m_{li}^{+}m_{jk}^{+}-m_{lj}^{+}m_{ik}^{+}\right\} \right]P\,,\nonumber \\
\widehat{\rho}\widehat{b}_{i}^{\dagger}\widehat{b}_{j}^{\dagger} & \longrightarrow & \left[m_{ij}^{+}-\frac{\partial}{\partial n_{lk}}\left\{ m_{li}^{+}n_{jk}-m_{lj}^{+}n_{ik}\right\} -\frac{\partial}{\partial m_{lk}}\left\{ n_{jl}n_{ik}-n_{il}n_{jk}\right\} -\frac{\partial}{\partial m_{lk}^{+}}\left\{ m_{li}^{+}m_{jk}^{+}-m_{lj}^{+}m_{ik}^{+}\right\} \right]P\,,\nonumber \\
\widehat{b}_{j}^{\dagger}\widehat{\rho}\widehat{b}_{i}^{\dagger} & \longrightarrow & \left[m_{ij}^{+}+\frac{\partial}{\partial n_{lk}}\left\{ m_{lj}^{+}n_{ik}+m_{li}^{+}\tilde{n}_{jk}\right\} -\frac{\partial}{\partial m_{lk}}\left\{ n_{il}\tilde{n}_{jk}-\tilde{n}_{jl}n_{ik}\right\} -\frac{\partial}{\partial m_{lk}^{+}}\left\{ m_{li}^{+}m_{jk}^{+}-m_{lj}^{+}m_{ik}^{+}\right\} \right]P\,,\label{eq:mn_correspondances}\end{eqnarray}

\twocolumngrid where we have used the Einstein summation convention
for repeated indices. Furthermore, we have explicitly written out
the extra derivative terms that arise from the antisymmetry of $\mathbf{m}$
and $\mathbf{m}^{+}$, such that the summation of these terms is only
for $k>l$. The factor of two difference between these equations and
the full covariance equations is due to the fact that these equations
correspond to a covariance which is constrained to satisfy the Hermitian
anti-symmetry condition.


\begin{thebibliography}{10}
\bibitem{everything}This form of Fermi gauge is also used in \cite{CorbozTroyer} to generate
phase-space equations for the general electronic structure problem.
\end{thebibliography}
\end{document}